\documentclass[useAMS,usenatbib]{mn2e}
\usepackage{psfig}

\title[The AGN Fraction and its Environment Dependence]
{The DEEP2 Galaxy Redshift Survey: The Red Sequence AGN Fraction and its
  Environment and Redshift Dependence} 

\author[Antonio D. Montero-Dorta et al.]{
\parbox[t]{\textwidth}{
Antonio D. Montero-Dorta$^1$, 
Darren J. Croton$^2$,
Renbin Yan$^{2,6}$,
Michael C. Cooper$^{2,5}$,
Jeffery A. Newman$^{2,7}$,
Antonis Georgakakis$^3$,
Francisco Prada$^1$, 
Marc Davis$^2$,
Kirpal Nandra$^3$,
Alison Coil$^4$
}
\vspace*{6pt} \\ 
$^1$Instituto de Astrof\'isica de Andaluc\'ia, Granada, Andaluc\'ia, Spain \\
$^2$Department of Astronomy, University of California, Berkeley, California, USA \\
$^3$Astrophysics Group, Imperial College London, Blackett Laboratory, London, United Kingdom\\
$^4$Hubble Fellow, Steward Observatory, University of Arizona, Tucson, Arizona, USA\\
$^5$Spitzer Fellow, Steward Observatory, University of Arizona, Tucson, Arizona, USA\\
$^6$Department of Astronomy and Astrophysics, University of Toronto, Canada\\
$^7$Department of Physics and Astronomy, University of Pittsburgh, Pittsburgh, PA, USA 
\vspace{-0.5cm} 
}

\date{Accepted ---. Received ---;in original form ---}

\pubyear{2005}

\newcommand{\plotone}[1]
           {\centering \leavevmode \psfig{file=#1,width=\columnwidth,clip=}}

\def\simlt{\lower.5ex\hbox{$\; \buildrel < \over \sim \;$}}
\def\simgt{\lower.5ex\hbox{$\; \buildrel > \over \sim \;$}}
\usepackage{graphicx}
\usepackage{rotating}

\begin{document}

\maketitle


\begin{abstract}

We measure the dependence of the AGN fraction on local environment at
$z\sim1$, using spectroscopic data taken from the DEEP2 Galaxy
Redshift Survey, and Chandra X-ray data from the All-Wavelength
Extended Groth Strip International Survey (AEGIS). To provide a clean
sample of AGN we restrict our analysis to the red sequence population;
this also reduces additional colour--environment correlations.  We
find evidence that high redshift LINERs in DEEP2 tend to favour higher
density environments relative to the red population from which they
are drawn.  In contrast, Seyferts and X-ray selected AGN at $z\sim1$
show little (or no) environmental dependencies within the same
underlying population. We compare these results with a sample of local
AGN drawn from the SDSS. Contrary to the high redshift behaviour, we
find that both LINERs and Seyferts in the SDSS show a slowly declining
red sequence AGN fraction towards high density environments.
Interestingly, at $z\sim1$ red sequence Seyferts and LINERs are
approximately equally abundant.  By $z\sim0$, however, the red Seyfert
population has declined relative to the LINER population by over a
factor of $7$. We speculate on possible interpretations of our
results. 
   
\end{abstract}


\section{Introduction}
\label{sec:intro}

Both active galactic nuclei (AGN) and local environment play key roles
in shaping galaxy evolution. It is now understood that AGN are those
nuclei in galaxies that emit radiation powered by accretion onto a
supermassive black hole. Although this realisation has proved useful
for explaining many observed characteristics of these active objects,
there are still many unsolved problems, especially related to the
physics of the accretion process itself.  In the recent years much
effort has been invested in studying the global properties of AGN as a
unique population in the context of galaxy formation. In this work, we
focus on a fundamental question: the dependence of the fraction of
galaxies that have AGN on the density of the local environment at
$z\sim1$, and the evolution of this dependence to $z\sim0$.

At low redshift, many authors have investigated various correlations
between \textit{galaxy properties} and environment. It is now well
established that there exists a relationship between morphology and
density (\citealt{Oemler1974} and \citealt{Dressler1980}), in that
star-forming disk-dominated galaxies tend to inhabit less dense
regions of the Universe than ``quiescent'' or inactive elliptical
galaxies.  Moreover, additional (and related) dependencies with
environment have been found, such as with stellar mass, luminosity,
colour, recent and past star formation, star formation quenching,
surface brightness, and concentration (to name but a few)
\citep[e.g.][]{Kauffmann2004, Balogh2004, Hogg2004, Blanton2005,
Bundy2006}.

In this scenario of entangled correlations it is useful to investigate
the dependence of AGN properties on the local environment, especially
since AGN are believed to play an important part in shaping galaxy
evolution. This has sometimes been a rather controversial issue. In
the local Universe, \cite{Miller2003} found no dependence on
environment of the fraction of spectroscopically selected AGN, using
the SDSS early data release. This result is in good agreement with
\cite{Sorrentino2006} who used the much larger SDSS DR4. However, many
other authors claim the existence of a strong link between nuclear
activity and environment, at least for specific AGN
types. \cite{Kauffmann2004} found that intermediate luminosity
optically selected AGN (Seyfert IIs) favoured underdense environments,
while low-luminosity optically selected AGN (Low-Ionization Nuclear
Emission-line Regions; hereafter, LINERs) showed no density
dependence, within the SDSS DR1. Similarly, lower-luminosity AGN were
found to have a higher clustering amplitude than high-luminosity AGN
by \cite{Wake2004} and \cite{Constantin2006a}.  Radio-loud AGN have
been noted to reside preferentially in mid-to-high density regions and
tend to avoid underdense environments \citep{Zirbel1997, Best2004}.

At high redshift the study of both galaxies and AGN, and their
relation to the environment, has been restricted by the lack of
adequate data. Only in recent years, with the emergence of quality
large-scale probes of the high redshift galaxy population, such as the
DEEP2 Galaxy Redshift Survey \citep{Davis2003} or the VIMOS-VLT Deep
Survey \cite[VVDS,][]{LeFevre2003}, have we reached the stage where we
can begin to measure the statistics of galaxy evolution in some
detail.  Using DEEP2, \cite{Cooper2006} found that the many of the low
redshift galaxy correlations with environment are already in place at
$z\!\sim\!1$. However important differences exist. The colour-density
relation, for instance, tends to weaken towards higher redshifts
\citep{Cooper2007a,Cucciati2006}.  Also, bright blue galaxies are
found, on average, in much denser regions than at low redshift. Such a
population inverts the local star formation-density relation in
overdense environments \citep{Cooper2007b,Elbaz2007}. This inversion
may be an early phase in a galaxy's transition onto the red sequence
through the process of star formation quenching. The truncation of
star formation in massive galaxies is believed to be tightly connected
with nuclear activity \citep[see e.g.][for more
information]{Croton2006, Bower2006}. Further investigation reveals
that post-starburst (aka. K+A or E+A) galaxies
\citep[e.g.][]{DresslerGunn1983} are galaxies ``caught in the act'' of
quenching and are in transit to the red sequence. These predominantly
``green valley'' objects reside in similar environments to regular
star forming galaxies (\citealt{Hogg2006, Nolan2007}; Yan et al. 2007
in prep.) supporting the picture that star formation precedes
AGN-triggered quenching, which precedes retirement onto the red
sequence.

\cite{Georgakakis2007} were one of the first to study the environments
of X-ray selected AGN at $z\sim1$ using a sample of 58 sources drawn
from the All-Wavelength Extended Groth Strip International Survey
(AEGIS, \citealt{Davis2007}). The authors found that these galaxies
avoided underdense regions with a high level of
confidence. \cite{Nandra2007} show that the same AGN reside in host
galaxies that populate from the top of the blue cloud to the red
sequence in colour-magnitude space.  They speculate that such AGN may
be the mechanism through which a galaxy stays red.  Similar ideas have
become a popular feature of many galaxy formation models that
implement lower luminosity (i.e. non-quasar) AGN to suppress the
supply of cooling gas to a galaxy, hence quenching star formation
through a process of ``starvation'' \citep[e.g.][]{Croton2006,
Bower2006}.

In this work we study the environmental dependence of nuclear activity
in red sequence galaxies within a carefully chosen sample of both
X-ray and optically selected AGN, drawn from the AEGIS Chandra
catalogue and the DEEP2 Galaxy Redshift Survey, respectively. Our
paper is organised as follows.  In Section~\ref{sec:survey} we
describe our AGN selection. In Section~\ref{sec:results} we present
our main result: the AGN fraction of red sequence galaxies at $z\sim1$
as a function of environment for three types of AGN (LINERs, Seyferts
and X-ray selected). We undertake a comparison between our high-z
results and those derived from a low-z sample drawn from the SDSS in
Section~\ref{sec:sdss}. Finally, in Sections~\ref{sec:discussion} and
\ref{sec:summary} we provide a discussion and brief summary.
Throughout, unless otherwise stated, we assume a standard $\Lambda$CDM
concordance cosmology, with $\Omega_m=0.3$, $\Omega_\Lambda=0.7$,
$w=-1$, and $h=1$. In addition, we use AB magnitudes unless otherwise
stated.

\begin{figure*}
\begin{center}
\begin{tabular}{c}
\includegraphics[scale=0.7]{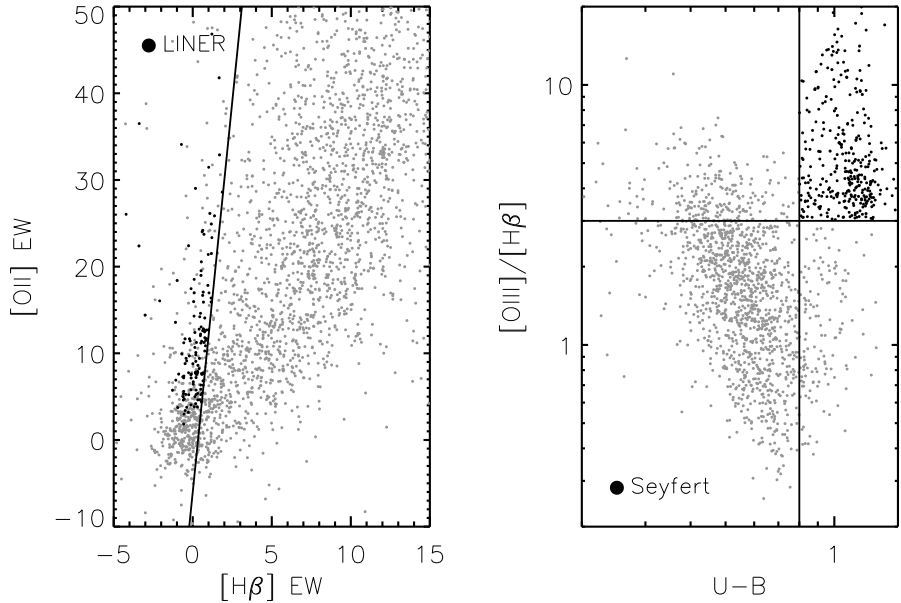}
\end{tabular}
\end{center}
\caption{Two panels that show our AGN selection of Seyferts and LINERS
within the DEEP2.  The left panel plots [OII] EW versus ${\rm
H{\beta}}$ EW for objects with accurate redshifts ($Q\ge3$),
$\delta_{3}$ environment measures, and covered [OII], [OIII] and ${\rm
H{\beta}}$ (grey points).  LINERs (black points) are selected using
the empirical demarcation of Equation~\ref{eqn:liners_selection} along
with the colour cut defined by Equation~\ref{eqn:color_deep2}.  The
right panel shows the line ratio ${\rm [OIII]/ H{\beta}}$ plotted
against $(U-B)$ rest-frame colour for the same DEEP2 sample (grey
points).  Seyferts (black points) are selected to have ${\rm
[OIII]/H{\beta}\ge 3}$ and rest-frame colour $(U\!-\!B)>0.8$, as
denoted by the horizontal and vertical lines.  See
Section~\ref{sec:optical} for further details.}
\label{fig:OSS_selection}
\end{figure*}

\section{Galaxy and AGN Selection}
\label{sec:survey}

Our primary galaxy and AGN samples are drawn from the DEEP2 Galaxy
Redshift Survey \citep{Davis2003,Davis2005}, a project designed to
study galaxy evolution and the underlying large-scale structure out to
redshifts of $z\sim1.4$. The survey utilises the DEIMOS spectrograph
\citep{Faber2003} on the 10-m Keck II telescope and has so far
targeted $\simgt 50\,000$ galaxies covering $\sim 3$ square degrees of
sky over four widely separated fields.  In each field, targeted
galaxies are observed down to an apparent magnitude limit of $R_{\rm
AB} < 24.1$.  Important for this work, the spectral resolution of the
DEIMOS spectrograph is quite high, ${\rm R} \approx 5000$, spanning an
observed wavelength range of $6500\!<\!\lambda\!<\!9200$\AA. This
allows us to confidently identify AGN candidates through emission-line
ratios down to low equivalent widths; such objects form a core part of
the data analysed in this paper.  More details on the DEEP2 survey
design and galaxy detection can be found in \cite{Davis2003,
Davis2005, Davis2007} and \cite{Coil2007b}.

To study the dependence of the AGN fraction on local environment, for
each galaxy we use the pre-calculated projected
third-nearest-neighbour distance, $D_{p,3}$, and surface density,
$\Sigma_{3}=3/(\pi D_{p,3}^{2})$ \citep[taken from][]{Cooper2005}.
This density measure is then normalised by dividing by the mean
projected surface density at the redshift of the galaxy in question,
yielding a quantity denoted by $1+\delta_3$ .  Tests using mock galaxy
catalogues show that $\delta_{3}$ is a robust environment measure that
minimises the role of redshift-space distortions and edge effects.
See \cite{Cooper2005} for further details and comparisons with other
commonly used density estimators.

To complement our optical catalogue we employ Chandra X-ray data from
the All-Wavelength Extended Groth Strip International Survey (AEGIS,
\citealt{Davis2007}).  The AEGIS catalogue provides a panchromatic
measure of the properties of galaxies in the Extended Groth Strip
(EGS) covering X-ray to radio wavelengths.  The EGS is part of DEEP2,
constituting approximately one sixth of its total area.  This allows
us to cross-correlate each X-ray detection with the optical catalogue
to identify each galaxy counterpart.  In this way environments can be
determined for the X-ray AGN sources.

Selecting objects from both the DEEP2 spectroscopic and Chandra
(AEGIS) X-ray catalogues provides two different AGN populations that
are embedded in the same underlying large-scale structure. To
differentiate the two in the remainder of the paper, we hereafter
refer to the first as the optically selected sample (OSS) and the
second as the X-ray selected sample (XSS). In the following sections
we describe the OSS and XSS populations in more detail.

\subsection{The optically selected AGN sample (OSS)}
\label{sec:optical}

Optically (or spectroscopically) selected AGN in the DEEP2 survey can
be divided into two main classes, LINERs and Seyferts, distinguished
primarily through the spectral lines present and their strength.
Although the physical processes that differentiate one class from the
other are still not well understood, the identification of each class
is never-the-less well defined.  We restrict our analysis to the
redshift range $0.72\!<\!z\!<\!0.85$ to ensure that all chosen AGN
spectral indicators are visible within the covered wavelength range
and that the environment measure is sufficiently reliable.  This will
be the redshift interval from which all our OSS results are
taken. Furthermore, to facilitate a fair comparison between both AGN
types, only objects on the red sequence or in the green valley are
included (defined below). This will also allow us to compare with a
low redshift sample (see Section~\ref{sec:sdss}).  For a complete
discussion of the spectroscopic detection of AGN in the DEEP2 survey
see Yan et al.  2007 (in prep.). Below we will briefly outline our
LINER and Seyfert selection in turn.

\subsubsection{LINERs}
As discussed in \cite{Yan2006}, LINERs are a population of
emission-line galaxies with high equivalent width (EW) ratio
${\rm[OII]/H{\alpha}}$ (or ${\rm[OII]/H{\beta}}$). Specifically, we
select a complete sample of LINERs using the division 
in ${\rm[OII]/H{\beta}}$ EW space given in \cite{Yan2006}:
\begin{equation}
{\rm EW \big(\big[O_{II}\big]\big)>18\,EW\big(H{\beta}\big)-6}
\label{eqn:liners_selection}
\end{equation}
The left panel of Figure~\ref{fig:OSS_selection} illustrates this
selection by plotting [OII] EW against ${\rm H{\beta}}$ EW for the
entire DEEP2 sample with accurate redshifts ($Q\ge3$), $\delta_{3}$
environment measures, covered [OII], [OIII] and ${\rm H{\beta}}$ (for
consistency with Seyfert selection -- see below), and redshift window
$0.72\!<\!z\!<\!0.85$ (grey points).  The solid line indicates the
empirical demarcation of Equation~\ref{eqn:liners_selection}.  Since
quiescent galaxies with no line emission also satisfy this criteria,
the inequality relation alone is not sufficient.  Thus, we further
require LINERs to have significant detection ($2\sigma$) of [OII]. As
${\rm H{\beta}}$ emission is expected to be weak in LINERs
\citep{Yan2006}, we do not require significant detection on ${\rm
H{\beta}}$.

The error on ${\rm H{\beta}}$ EW emission is large due to the
difficulty in measuring it after subtracting the stellar absorption.
The above LINER selection has contamination from star-forming galaxies
whose ${\rm H{\beta}}$ EW is underestimated. From a study of SDSS
galaxies, \cite{Yan2006} concluded that LINERs are almost exclusively
found in red sequence galaxies.  Therefore, we adopt an additional
colour cut to remove this contamination, which is the same used by 
\cite{Willmer2006}:
\begin{equation}
{(U\!-\!B)>-0.032M_{B}+0.322}
\label{eqn:color_deep2}
\end{equation} 
Our final LINER sub-sample with all of the above constraints is
comprised of 116 objects and is over-plotted in the left panel of
Figure~\ref{fig:OSS_selection} with black points.  Note that within
the SDSS a strong vertical branch can be seen (see Figure~2 of
\citealt{Yan2006}, where they use $\rm H{\alpha}$ instead of ${\rm
H{\beta}}$).  This branch is significantly weaker at $z\sim1$ in the
DEEP2 data. This is due in part to the greater errors on ${\rm
H{\beta}}$ in the DEEP2 data, and in part to the domination of red
galaxies in the SDSS sample (due to the SDSS selection criteria).

\begin{figure}
\plotone{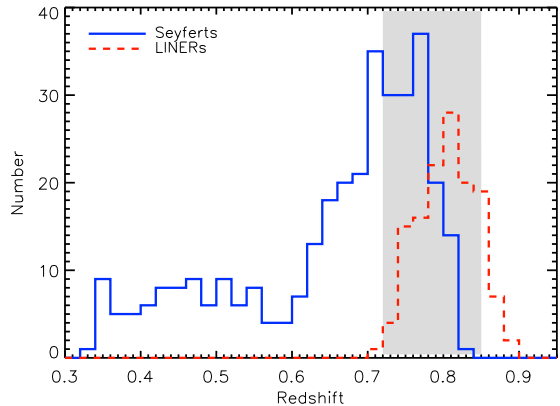}
\caption{The redshift distribution for our optically selected AGN
sample (OSS, Section~\ref{sec:optical}). The distribution of Seyferts
is given by the blue solid line, while the distribution of LINERs is
given by the red dashed line. The shaded region denotes the redshift
window $0.72\!<\!z\!<\!0.85$ from which our final OSS sample is drawn.
Within this window both populations are cleanly identified
spectroscopically and the effect of selection is small in both
sub-samples. }
\label{fig:OSS_redshift}
\end{figure}

\subsubsection{Seyferts}
Seyferts require different selection techniques than LINERs.
Following the method of Yan et al (in prep.), we identify Seyferts in
DEEP2 using a modified Baldwin-Phillips-Terlevich (BPT) diagram
\citep{Baldwin1981}. Historically, the BPT diagram has been a reliable
tool for determining the source of line emission from a galaxy. By
plotting the line ratios ${\rm [OIII]\ \lambda 5007 / H{\beta}}$
against ${\rm [NII]\ \lambda 6583/H{\alpha}}$ one can visually
differentiate Seyferts, LINERs and star-forming galaxies.  However,
${\rm H{\alpha}}$ is not available in the DEEP2 spectra at
$z\simgt0.4$ as it is redshifted into the infrared. For this reason,
we use a modified BPT diagram which replaces the line ratio ${\rm
[NII]\ \lambda 6583/H{\alpha}}$ with the rest-frame $U\!-\!B$ colour.
This is possible because both are rough proxies for metallicity. Tests
done on SDSS samples demonstrate that such a substitution is able to
produce a clean and complete selection criterion for Seyferts (Yan et
al. in prep.).

In the right panel of Figure~\ref{fig:OSS_selection} we illustrate our
Seyfert selection by showing the modified BPT diagram for the same
underlying sample used to select LINERs (grey points).  This figure
shows that the modified BPT diagram has a similar two branching
structure to the original BPT diagram. Seyferts are selected to have
${\rm [OIII]/H{\beta}\ge 3}$ and rest-frame colour $(U\!-\!B)>0.8$
(horizontal and vertical lines, respectively).  For cases in which
${\rm H{\beta}}$ is not positively detected, we use a $2\sigma$ lower
limit on ${\rm [OIII]/H{\beta}}$. With such criteria we obtain 131
Seyferts in the range $0.72\!<\!z\!<\!0.85$ where all spectral
signatures for both Seyferts and LINERs are normally available (black
points). Selecting only red sequence (or green valley) objects
facilitates a fair comparison with LINERs and is consistent with their
typical position in the colour-magnitude diagram \citep{Yan2006}.

\begin{figure}
\plotone{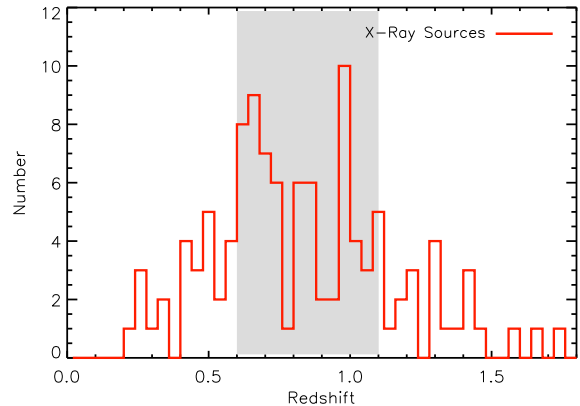}
\caption{The redshift distribution for our X-ray selected AGN sample
(XSS, Section~\ref{sec:xray}), shown by the red solid line. The grey
shaded region denotes the redshift range $0.6\!<\!z\!<\!1.1$, from
which our final XSS is drawn.  This range was chosen to be
approximately comparable to the OSS (Figure~\ref{fig:OSS_redshift})
while simultaneously maximising the number and completeness in the
sample.}
\label{fig:XSS_selection}
\end{figure}

\subsubsection{Redshift distributions}

In Figure~\ref{fig:OSS_redshift} we show the redshift distribution of
both LINERs (dashed line) and Seyferts (dotted line) drawn from the
selection given in each panel of Figure~\ref{fig:OSS_selection}. The
DEEP2 Seyfert population extends from $z \approx 0.35$ to $z \approx
0.85$, peaking at around $0.75$. For LINERs the distribution is much
more concentrated, extending from $0.72$ to $0.9$ and peaking at
around $0.8$. Note that the peak for both is dominated by the DEEP2
survey galaxy selection and not an intrinsic peak in the AGN
distribution.
As discussed previously, the redshift window where both populations
can be cleanly identified spectroscopically is $0.72\!<\!z\!<\!0.85$,
denoted by the shaded region. This range maximises AGN coverage while
insuring that selection effects are minimised in both samples.  

\begin{figure*}
\begin{center}
\begin{tabular}{c}
\includegraphics[scale=0.7]{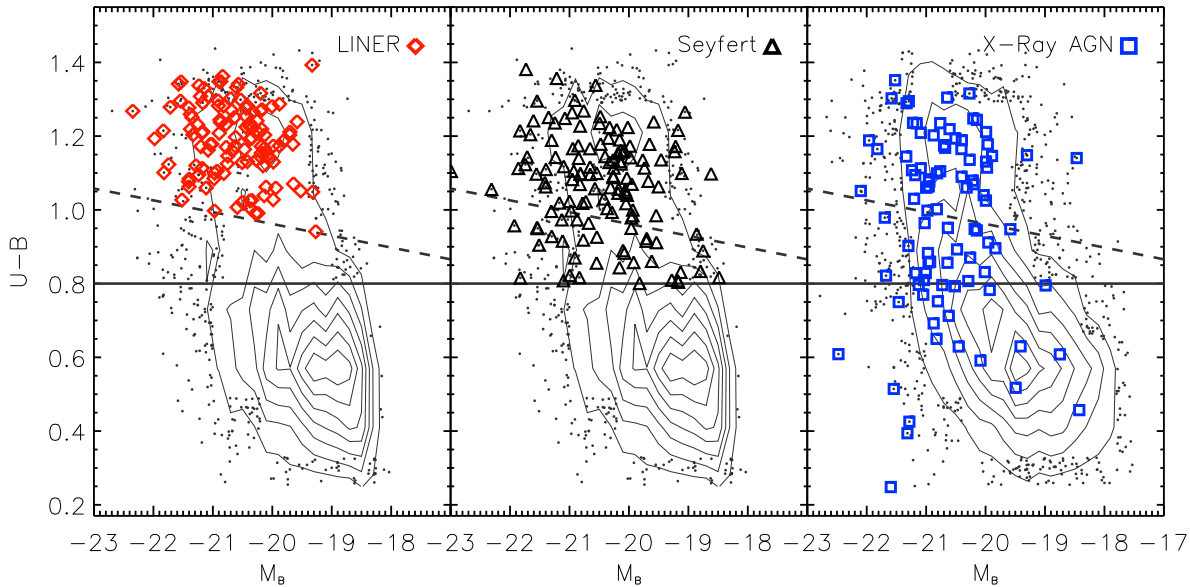}
\end{tabular}
\end{center}
\caption{The colour-magnitude diagram (CMD) for LINERs (left panel, red
diamonds), Seyferts (middle panel, black triangles) and X-ray AGN
(right panel, blue squares). The demarcations given by the solid and
dashed lines represent the conventions adopted to separate the blue
cloud from the green valley, and the latter from red sequence objects
(Equation~\ref{eqn:color_deep2}), respectively \citep{Willmer2006}. The
LINER sub-sample is composed of 116 objects, all of them lying on the
red sequence by definition. The Seyferts sub-sample is composed of 131
objects, with 97 of them on the red sequence and 34 in the green
valley. Finally, from our X-ray sample of 68 objects, 36 sources are
red, 16 are green, and the remaining 16 blue.  The underlying CMD of
the population from which all AGN are drawn is shown in each panel
with grey contours and black points. This parent population, 
in the left-hand and middle panels,
 is comprised of objects with accurate redshifts ($Q\ge3$), 
$\delta_{3}$ environment measures, covered [OII], [OIII] and ${H{\beta}}$
; and redshift between $0.72$ and $0.85$. In the right-hand panel,
the grey contours and black points represent all objects in the EGS field with  
accurate redshifts ($Q\ge3$) and  $\delta_{3}$ environment measures;
and redshift between $0.6$ and $1.1$.}
\label{fig:CMD}
\end{figure*}

\subsection{The X-Ray selected AGN sample (XSS)}
\label{sec:xray}

AEGIS Chandra X-ray sources within the EGS field are optically and
spectroscopically identified by cross-correlating with the DEEP2
photometric and redshift catalogues, following the prescriptions
presented by \cite{Georgakakis2007}. They cover X-ray luminosities of
$10^{41}\!\simlt\!{\rm L_X (erg/s)}\!\simlt\!10^{44}$ in host galaxies
of luminosity $-19\!\simlt\!M_{B}-5\log h\!\simlt\!-22$. The base
X-ray sample comprises a total of $113$ reliably matched objects.

In Figure~\ref{fig:XSS_selection} we show the redshift histogram of
our X-ray catalogue.  To extract a sample that is as closely
comparable to the OSS as possible while simultaneously maximising AGN
number and completeness, we restrict the X-ray sources to the redshift
range $0.6\!<\!z\!<\!1.1$.  This is wider than the OSS redshift
window; however, both samples (OSS and XSS) have similar redshift
means, and we assume that the evolution effects for sources outside
the OSS redshift window do not dominate our results (or at least does
not differ significantly from evolution in the red sequence population
itself).  In this redshift range the number of reliable X-ray AGN
drops to $68$, including $52$ red-ward of $(U\!-\!B)>0.8$ (i.e. a green
valley cut), and $36$ red-ward of Equation~\ref{eqn:color_deep2}
(i.e. a red sequence cut).

\subsection{AGN in colour-magnitude space}
\label{sec:cmd}

In Figure~\ref{fig:CMD} we show the colour-magnitude diagram (CMD) for
LINERs (left panel), Seyferts (middle panel) and X-ray AGN (right
panel). The demarcations given by the solid and dashed lines represent
the conventions adopted to separate the blue cloud from the green
valley, and the latter from red sequence objects
(Equation~\ref{eqn:color_deep2}), respectively (\citealt{Willmer2006};
Yan et al. in prep.).  Here, LINERs are red sequence galaxies by
definition. As explain above, this restriction is supported by the
fact that local LINERs are almost exclusively red \citep{Yan2006}. For
Seyferts, $\sim 80\%$ lie on the red side of the CMD, with the
remainder residing in the green valley. Finally, for the XSS AGN,
$\sim 50 \%$ of the sources are red, $\sim 25 \%$ are green, and the
remaining $\sim 25 \%$ blue.  The grey contours in each panel show the
underlying DEEP2 CMD within the same redshift range.

\begin{figure*}
\begin{center}
\begin{tabular}{c}
\includegraphics[scale=0.7]{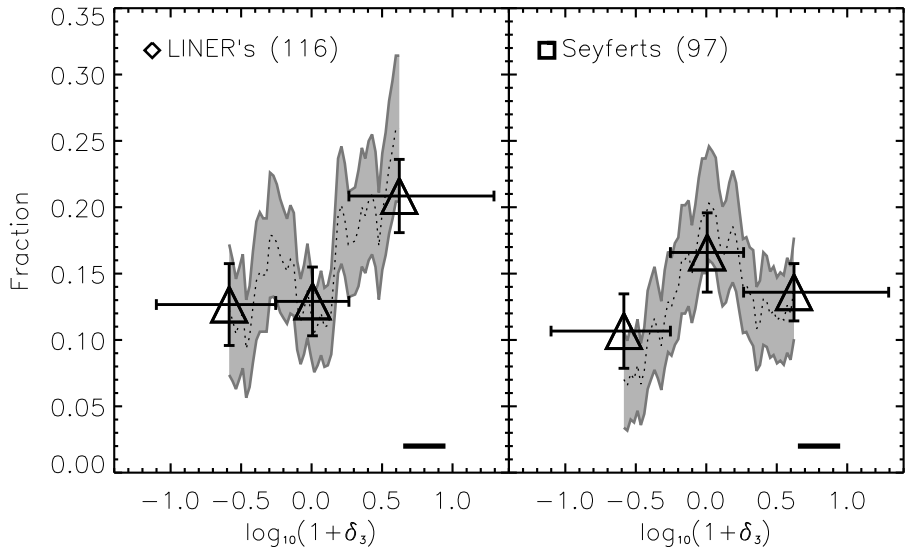}
\end{tabular}
\end{center}
\caption{The AGN fraction in red sequence galaxies versus local galaxy
over-density, $\delta_{3}$, for LINERs (left panel) and Seyferts
(right panel). For each, the respective symbols (diamonds for LINERs
and squares for Seyferts) show the median measure in bins of low, mean
and high density.  Vertical error-bars represent the Poissonian
uncertainty, while horizontal error-bars show the size of each density
range. We also show how the AGN fraction varies smoothly with
environment using a sliding box of width $0.3$ dex shifted from low to
high density in increments of $0.025$ dex (dotted lines with shaded
area showing the $1\sigma$ uncertainty in the sliding fraction). The
overall fraction of LINERs and Seyferts is plotted with horizontal
dashed lines.  This figure shows some evidence that LINERs tend to
favour high density environments relative to the underlying red
sequence, whereas Seyferts have little (or no) environment
dependencies.}
\label{fig:OSS_result}
\end{figure*}

\subsection{Errors and completeness}
\label{sec:errors}

Our greatest source of error is that of noise from small number
statistics, given the low number of AGN we have available in the DEEP2
and AEGIS surveys in any particular environment bin.  Such current
generation high redshift catalogues are thus still limited in the
extent to which the statistical nature of the AGN population can be
examined. All errors calculated in this paper were determined by
propagating the Poissonian uncertainties on the number of objects. Due
to the small number statistics, the errors obtained with this method
will dominate any cosmic variance effects in the observed fields
\citep{NewmanDavis2002}.

It should be noted that the DEEP2 survey is by design incomplete. At
$z\sim1$, approximately $60\%$ of the actual objects are observed by
the telescope. Moreover, redshifts are successfully obtained for around 
$70\%$ of the target parent population (based on tests with blue 
spectroscopy, most failures are objects at z>1.4 (Steidel, priv. comm.)). 
This should be carefully considered in any statistic that counts 
absolute numbers of objects.  In our work, however, we deal with 
\emph{relative} numbers of objects, i.e.  the
AGN fraction.  We assume, to first order (and to the level of
uncertainty given by the Poisson error), that any variation in
redshift success or targeting rate between the AGN sample and the red
sequence parent population is the same in low density regions as it is
in high density regions.  In principle, one may expect an easier
detection (or even an easier redshift estimation) of an object
identified as an AGN than the one for a ``regular'' red sequence
object (due to the presence of remarkable features in the spectrum).
However both \cite{Cooper2005} and \cite{Gerke2005} found that DEEP2
selection rates are essentially independent of local density.

Finally, because of the different Seyfert and LINER selection we find
some inevitable (but small) overlap between the two populations, $7\%$
of the total in our case, where a single object has been classified as
both AGN types. We have re-calculated all our results excluding these
dual class objects and find only trivial differences.  For the sake of
maximising statistics we have not removed such objects from the OSS,
however note that they may constitute an interesting sub-population
whose physical implications warrant further investigation.

\section{Results}
\label{sec:results}

In this section we present our primary result: the dependence of the
AGN fraction in the red sequence on local environment density.  We
will also extend the analysis to include green valley objects.

Figure~\ref{fig:OSS_result} presents the density dependence of the
fraction of $z\sim1$ red sequence AGN, for LINERs (left panel) and
Seyferts (right panel) separately. In each panel, the respective
symbols show the median measure in bins of low, mean and high density
environments (each of them encompassing one third of the OSS), where
the horizontal error-bars indicate the width of each bin, and the
vertical error-bars show the Poisson uncertainty in the measured
fraction, as described in Section~\ref{sec:errors}.  We also show how
the AGN fraction varies smoothly with environment using a sliding box
of width $0.3$ dex, shifted from low to high density in intervals of
$0.025$ dex (dotted line). The accompanying grey-shaded regions
correspond to the sliding $1\sigma$ uncertainties in the sliding
fraction.

\begin{figure*}
\begin{center}
\begin{tabular}{c}
\includegraphics[scale=0.7]{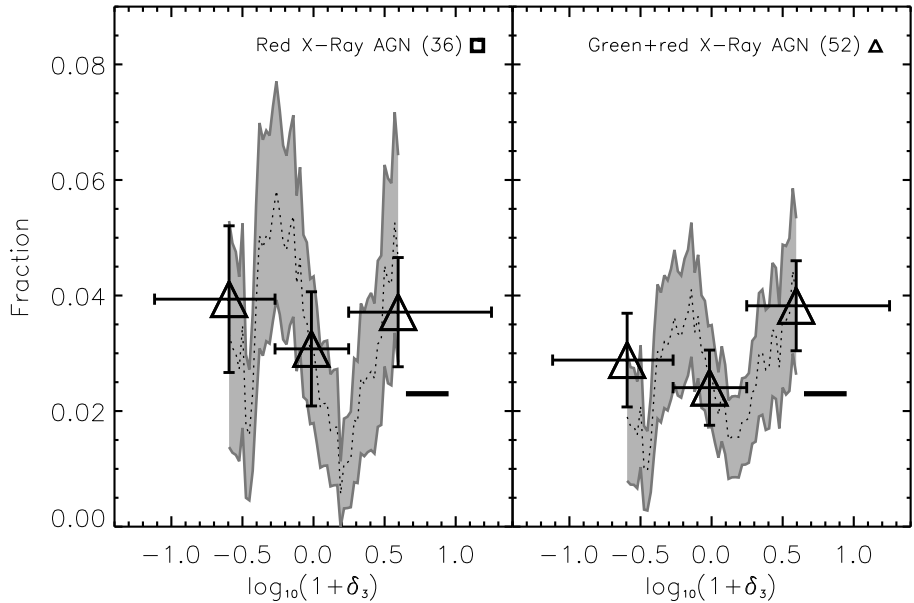}
\end{tabular}
\end{center}
\caption{The AGN fraction versus environment for our X-ray selected
sample (XSS), using the same format as in
Figure~\ref{fig:OSS_result}. In the left panel, squares show the
fraction of red sequence X-ray AGN in the three density bins
considered. In the right panel we extend this analysis to include
green valley AGN. Note that this extension does not change the results
in any significant way.  XSS AGN behave more like Seyferts than LINERs
(see Figure~\ref{fig:OSS_result}), with the fraction showing only a
weak (or no) environmental dependence to within the errors.}
\label{fig:XSS_result}
\end{figure*}

Evidence for a trend in the behaviour of the LINERs is quite apparent
in Figure~\ref{fig:OSS_result}, suggesting the possibility that these
objects tend to favour high density environments, and in a way
stronger than the majority of red sequence galaxies. This is in
contrast to the behaviour of red Seyferts, which show little (or
no) environment dependence relative to the red sequence.  This is
a key result that will be discussed in more detail in the following
sections.

We now consider the X-ray catalogue drawn from the AEGIS Chandra
imaging.  Figure~\ref{fig:XSS_result} presents the X-ray selected AGN
fraction versus local environment (note that the same format used in
Figure~\ref{fig:OSS_result} has been applied here).  In the left panel
we show the fraction of red X-Ray AGN and in the right panel we
extended the analysis to include green valley X-ray AGN and galaxies.
Including green valley objects does not significantly alter our
results.  Red sequence X-ray selected AGN appear to behave similarly
to optically selected Seyferts in terms of their lack of environmental
preference, and differently from the LINER population in high density
environments.  This is in agreement with the results of
\cite{Georgakakis2007}, also using the AEGIS data.  We note that
improved statistics could show trends at levels equal to or smaller
than that which can be measured here given our errors.

We have tested the significance of the results in
Figures~\ref{fig:OSS_result} and \ref{fig:XSS_result} in a number of
ways.  Since LINERs show the most interesting environment dependencies
we will focus our tests on this population.  We randomly draw $1000$
sub-samples from the red sequence and replace the LINER sample with
each of these random populations.  After repeating our analysis for
each we find that only $2\%$ of the random sub-samples show similar
density dependencies to the LINER population (i.e. results at least as
pronounced as the one in Figure~\ref{fig:OSS_result}) . The LINER
environment dependence seen in the left panel of
Figure~\ref{fig:OSS_result} deviates by at least $2\sigma$ (actually
almost $2.5\sigma$) from a random selection of red sequence galaxies.
Additionally, we can confirm that the trend in
Figure~\ref{fig:OSS_result} is not due to an implicit dependence of
colour or magnitude on environment within the red sequence.  This was
checked by repeatedly replacing the LINER sample with randomly drawn
objects with the same colour or colour and magnitude distributions,
and comparing their density distributions with that of the real
LINERs. The mean density of the LINER population is $1+\delta_3 = 0.37
\pm 0.06$, almost double that for randomly colour selected samples
which have $1+\delta_3 = 0.20 \pm 0.01$, and randomly colour and
magnitude selected samples with $1+\delta_3 = 0.19 \pm 0.01$.  Similar
tests were performed on Seyferts and X-Ray AGN.

\begin{figure*}
\begin{center}
\begin{tabular}{c}
\includegraphics[scale=0.7]{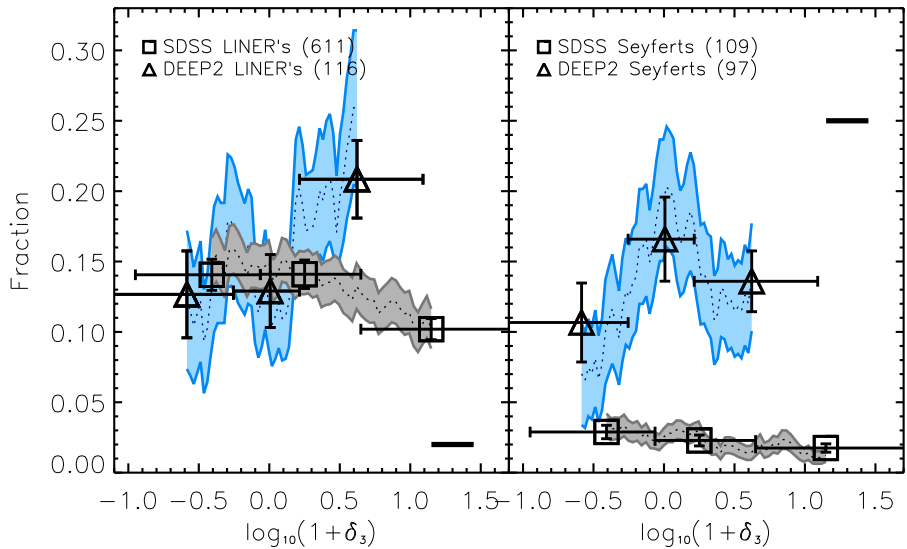}
\end{tabular}
\end{center}
\caption{The SDSS red sequence AGN fractions versus environment,
plotted using the same format from Figures~\ref{fig:OSS_result} and
\ref{fig:XSS_result}.  The left panel shows the fraction of low-z SDSS
LINERs (squares) and, for comparison, the high-z DEEP2 LINER fraction
(triangles) reproduced from Figure~\ref{fig:OSS_result}. The right
panel shows the fraction of SDSS Seyferts, with the equivalent DEEP2
result again reproduced from Figure~\ref{fig:OSS_result} for
comparison.  SDSS LINERs and Seyferts both show a decreasing AGN
fraction towards high density environments, unlike that seen in DEEP2.
At $z\sim1$, LINERs and Seyferts are approximately equally abundant,
whereas by $z\sim0$ the relative abundance of Seyferts to LINERs has
dropped by approximately a factor of $7$.}
\label{fig:SDSS_result}
\end{figure*}

\section{A Comparison with Local AGN in the SDSS}
\label{sec:sdss}

Our results thus far suggest that the $z\sim1$ red sequence LINER
fraction depends on environment in a way that is different from
Seyferts.  This dependence takes the form of an increase in the
relative abundance of LINERs in higher density environments.  In this
section we address the question of AGN fraction
evolution. Specifically, do local red sequence LINERs also favour
dense environments and Seyferts show little environment dependence?
 
Our low redshift AGN sample is drawn from the Sloan Digital Sky Survey
(SDSS, \citealt{York2000}) spectroscopic DR4 catalogue
\citep{Adelman2006}.  The SDSS DR4 covers almost $5000$ square degrees
of the sky in five filters ($ugriz$) to an apparent magnitude limit of
$r=17.7$.  The redshift depth is approximately $z\sim0.3$, with a
median redshift of $z=0.1$.  DR4 consists of $\sim400\,000$ galaxies.
The same environment measure is applied for consistency with the DEEP2
analysis above (see \citealt{Cooper2007a} for full details).

To measure the low redshift AGN fraction we follow a similar procedure
to that used for the high-z results. This procedure isolates a well
defined red sequence population and identifies AGN within it.  The
base red sequence population is constructed by selecting SDSS galaxies
within the redshift interval $0.05\!<\!z\!<\!0.15$ and applying the
rest-frame colour cut $(U-B)> -0.032M_{B}+0.483$
(\citealt{Cooper2007a}, in agreement with the previous analysis by
\citealt{Blanton2006}).  For consistency with our DEEP2 sample we take
a faint absolute magnitude limit of $M_{\rm B} - 5\log h = -20$
(representing the approximate faint-end of the red sequence at
$z\sim0.8$ within the DEEP2 data) and evolve it $0.88$ magnitudes to
mimic the evolution in the galaxy luminosity function between DEEP2
and SDSS (assuming evolution of $1.3$ magnitudes per unit redshift
from \citealt{Willmer2005} and mean DEEP2 and SDSS redshifts of $0.78$
and $0.1$, respectively).  With these constraints the underlying low-z
red galaxy sample is composed of $5335$ objects.

To select AGN from the SDSS red sequence sample we use the same set of
criteria described in Section \ref{sec:optical} with the following
modifications. Since SDSS spectra have a much higher signal-to-noise
than DEEP2 spectra the emission line detections are easier. This will
result in differences between the two AGN samples, as SDSS AGN will
include weaker optical AGN than the DEEP2 can detect. Therefore, we
determine a different line detection criterion by comparing the errors
in the emission line EW measurements. Typical line measurement errors
in DEEP2 are almost exactly twice as large as those in SDSS.  We thus
change all $2\sigma$ line detection criteria to $4\sigma$ for
selecting AGN in the SDSS.  The final low redshift AGN sample is
comprised of $720$ objects, $611$ being LINERs and $109$ Seyferts.
This should be contrasted with the high-z sample which has $213$
objects, of which $116$ are LINERs and $97$ Seyferts.

In Figure~\ref{fig:SDSS_result} we present the SDSS red sequence AGN
fractions versus environment (this figure follows the same format used
in Figures~\ref{fig:OSS_result} and \ref{fig:XSS_result}).  The left
panel shows the fraction of local red sequence LINERs (squares) and,
for comparison, the high redshift LINER fraction (triangles)
reproduced from Figure~\ref{fig:OSS_result}. The right panel shows the
Seyfert fraction in the SDSS red population (squares) and similarly
for DEEP2 (triangles, from Figure~\ref{fig:OSS_result}).

The left panel of Figure~\ref{fig:SDSS_result} reveals a different
LINER trend with environment at $z\sim0$ than that seen at $z\sim1$.
LINERs in the SDSS show no indication of favouring high density
regions relative to other environments. In fact, it is very
statistically significant that SDSS LINERs tend to reside more in
mean-to-low density environments and clearly disfavour those of high
density.  In the right panel of Figure~\ref{fig:SDSS_result} Seyferts
also follow a clear trend of decreasing AGN fraction towards denser
SDSS environments.  This is in contrast to the weak (or no)
environmental trend in the high redshift DEEP2 Seyfert population.

It is important to note that comparing the overall amplitudes of the
LINER and Seyfert fractions between DEEP2 and SDSS is dangerous, as
subtleties in the selection of the underlying red sequence can shift
the absolute values around somewhat.  The relative trends across
environment \emph{within} a population are much more robust however,
and these can be compared between high and low redshift (for example,
we do not expect any selection effects to have a significant density
dependence).  Additionally, the difference in the abundance
\emph{between} Seyferts and LINERs at a given redshift can be
contrasted.  Seyferts and LINERs are approximately equally abundant at
$z\sim1$.  By $z\sim0$ however, the Seyfert population has diminished
relative to the LINER population by over a factor of $7$.  This
decline in the relative number of Seyfert AGN by redshift zero will be
discussed in the next section.

\section{Discussion}
\label{sec:discussion}

\subsection{Previous measures of AGN and environment}

It is difficult to make direct comparisons of our results with
previously published works.  This is because past environment studies
have tended to focus on the AGN fraction of \emph{all} colours of host
galaxies, and also to mix both LINER and Seyfert classes into a
combined AGN population.  Our selection is restricted to the red
sequence only (and also green valley), which allows us to compare high
and low redshift populations and also study the AGN--environment
connection without the colour--environment correlation.

Locally, a number of SDSS measures of AGN and environment have been
made.  Using the SDSS early data release, \cite{Miller2003} found no
dependence on environment for the spectroscopically selected AGN
fraction in a sample of 4921 objects. Specifically, the authors report
no statistically significant decrease in the AGN fraction in the
densest regions, although their densest points visually suggest such a
trend. This result is broadly consistent with the results of both
LINERs and Seyferts in Figure~\ref{fig:SDSS_result}, even though we
only consider red sequence objects.

\cite{Kauffmann2003} also found little environment dependence of the
overall fraction of detected AGN in a sample drawn from the SDSS
DR1. However, they do report different behaviour when the sample is
broken into strong AGN ($\log L[{\rm OIII}]>7$, ``Seyferts'') and weak
AGN ($\log L[{\rm OIII}]<7$, ``LINERs''). For Seyfert they find a
significant preference for low-density environments, especially when
hosted by more massive galaxies. This is consistent with our SDSS
findings in Figure~\ref{fig:SDSS_result} and different to what we find
at $z\sim1$ in the DEEP2 fields. For LINERs,
\citeauthor{Kauffmann2003} measure little environment dependence,
whereas we find a significant decline in the SDSS LINER fraction in
our overdense bins.  The explanation for this difference may come from
our removal of possible contaminating star forming galaxies by
restricting our analysis to the red sequence. Also, we impose a higher
line detection threshold on the SDSS data to provide a fair comparison
with DEEP2 (see Section~\ref{sec:sdss}). Finally, \cite{Kauffmann2003}
required that all lines for the BPT diagnosis were detected and this
implies biasing the LINERs sample towards the strongest objects.

Between redshifts $z=0.4$ and $z=1.35$, \cite{Cooper2007a} show that
\emph{red galaxies} within the DEEP2 survey favour overdense
environments, although the blue fraction in clusters does become
larger as one moves to higher redshift \citep[see also][]{Gerke2007,
Coil2007b}.  At all redshifts there exists a non-negligible red
fraction in underdense environments, which evolves only weakly if at
all.  \cite{Nandra2007} show that the host DEEP2 galaxies of X-ray
selected AGN within the EGS field ($\sim 1/6$ of the DEEP2 survey
volume) occupy a unique region of colour-magnitude space.  These
objects typically live at the top of the blue cloud, within the green
valley, or on the red sequence.  \cite{Georgakakis2007} measure the
mean environment of this population and confirm that, on average, they
live in density regions above that of the mean of the survey.  They
find this to be true for all host galaxy magnitudes studied
($M_B\simlt-21$) and colours ($U-B\simgt0.8$) (note the DEEP2 red
sequence begins at $U-B\sim1$).  However, given limited sample sizes,
they were not able to establish whether the environment distribution
of the X-ray AGN differed from that of the red population, rather than
the DEEP2 population as a whole.

\subsection{Understanding the sequence of events}

From our results alone a comprehensive understanding of the different
environment trends within the AGN population from high to low redshift
is not possible.  However, some speculation and interpretation can be
made by drawing on our broader knowledge of these active objects from
the literature.  

One possible scenario posits that LINERs and Seyferts occur in
different types of galaxies. In this picture, LINERs are often
associated with young red sequence galaxies \citep{Graves2007} and are
especially common among post-starburst (K+A) galaxies
\citep{Yan2006}. These galaxies would already be into the quenched
phase of their evolution but still relatively young.  Merger triggered
starbursts and subsequent quasar winds are a possible mechanisms to
produce rapid star formation shut down in such objects
\citep{Hopkins2006}. The gas rich merging events required in this
scenario are common in overdense environments at $z\sim1$ as clusters
and massive groups assemble.  By $z\sim0$, however, the activity in
these environments has mostly ended.  Hence, if this picture is
correct, one may expect an over-abundance of red sequence LINERs in
dense environments at high redshift (since both star formation and
rapid quenching is common) that is not seen locally.  This may be
consistent with the trends found in the left panel of
Figure~\ref{fig:SDSS_result}.

Seyfert galaxies, on the other hand, could be objects in transition
from the blue cloud to red sequence \citep{Groves2006}, whose AGN are
thought to be initiated by internal processes (and not mergers),
inferred from their often found spiral structure (e.g. M77) (mergers
act to destroy such structure).  From this, one may expect our red
sequence Seyfert population to represent the tail of the colour
distribution of transitioning objects whose dependence on environment
is determined by secular mechanisms and who would evolve accordingly.
At high redshift, disk galaxies are commonly found in all
environments, including the most dense.  In contrast, overdense
regions in the local Universe are dominated by passive ellipticals and
show an absence of spirals.  This would be broadly consistent with our
findings in the right panel of Figure~\ref{fig:SDSS_result}, where the
most significant evolution in the red Seyfert fraction arises from a
depletion in overdense regions relative to other environments, from
high redshift to low.

Alternatively, some authors claim that LINERs and Seyferts form a
continuous sequence, with the Eddington rate the primary
distinguishing factor \citep{Kewley2006}. In this scenario, Seyferts
are young objects with actively accreting black holes. As the star
formation begins to decay so does the accretion rate, and the galaxy
enters a transition phase. Eventually, a LINER-like object emerges,
with an old stellar population and very low supermassive BH accretion
rate. This picture is supported by recent studies in voids from
\cite{Constantin2007}. At high redshift, our results show that red
Seyferts and LINERs are approximately equally abundant. By $z\sim0$
however, the Seyfert population has declined relative to the LINER
population by over a factor of $7$. This may be interpreted as the
natural transformation of Seyferts into LINERs with time, within a
galaxy population which is smoothly reddening from $z\sim1$ to
$z\sim0$.  Moreover, the fact that high-z LINERs reside preferentially
in high-density environments may imply that this Seyfert-LINER
transition is more efficient in dense regions of the Universe.

\section{Summary}
\label{sec:summary} 

In this paper we measure the dependence of the AGN fraction of red
galaxies on environment in the $z\sim1$ DEEP2 Galaxy Redshift Survey
and local $z\sim0.1$ SDSS.  We restrict our analysis to the red
sequence to maintain a clean and consistent selection of AGN at high
and low redshift, and this also reduces the additional effects of
environment associated with galaxy colour.

Our results can be summarised as follows:
\begin{itemize}
\item[(i)] High redshift LINERs at $z\sim1$ in DEEP2 appear to favour 
higher density environments relative to the red sequence from which
they are drawn.  In contrast, Seyferts and X-ray selected AGN at
$z\sim1$ show much weaker (or no) environmental dependencies within
the same underlying population.  Extending our analysis to include
green valley objects has little effect on the results.
\item[(ii)] Low redshift LINER and Seyfert AGN in the SDSS both show
a slowly declining red sequence AGN fraction towards high density
environments.  This is in contrast to the high redshift result.
\item[(iii)] At $z\sim1$, Seyferts and LINERs are approximately
equally abundant.  By $z\sim0$ however, the Seyfert population has
declined relative to the LINER population by over a factor of $7$.
\end{itemize}
    
It is important to remember that such measures are difficult to make
with current data, and hence we remain limited by statistics to the
extent to which we can physically interpret our results.  Regardless,
a robust outcome of our analysis is the differences between LINER and
Seyfert AGN populations in high density regions, and between high and
low redshift in all environments.  Our results indicate that a greater
understanding of both AGN and galaxy evolution may be possible if
future analyses simultaneously focus on the detailed subdivision of
different AGN classes, host galaxy properties, and their environment.

\section*{Acknowledgements}

AMD is supported by the Ministerio de Educaci\'on y Ciencia of the
Spanish Government through FPI grant AYA2005-07789, and wishes to
thank the University of California Berkeley Astronomy Department for
their hospitality during the creation of this work. DC acknowledges
support from NSF grant AST-0071048. ALC is supported by NASA through
Hubble Fellowship grant HF-01182.01-A, awarded by the Space Telescope 
Science Institute, which is operated by the Association of Universities
for Research in Astronomy, Inc., for NASA, under contract NAS 5-26555.
Support for this work was provided by NASA through the Spitzer Space 
Telescope Fellowship Program.

Funding for the DEEP2 survey has been provided by NSF grants
AST-0071048, AST-0071198, AST-0507428, and AST-0507483.  The data was
obtained at the W. M. Keck Observatory, which is operated as a
scientific partnership among the University of California, Caltech and
NASA. The Observatory was made possible by the generous financial
support of the W. M. Keck Foundation. The DEEP2 team and Keck
Observatory acknowledge the very significant cultural role and
reverence that the summit of Mauna Kea has always had within the
indigenous Hawaiian community and appreciate the opportunity to
conduct observations from this mountain.  The DEEP2 and AEGIS websites
are http://deep.berkeley.edu/ and http://aegis.ucolick.org/.

Funding for the SDSS has been provided by the Alfred P. Sloan
Foundation, the Participating Institutions, NASA, the NSF, the
U.S. Department of Energy, the Japanese Monbukagakusho, and the Max
Planck Society. The SDSS website is http://www.sdss.org/.

\bibliographystyle{mnras}
\bibliography{./paper}

\label{lastpage}

\end{document}